\documentclass[lettersize,journal]{IEEEtran}
\usepackage{amsmath,amsfonts}
\usepackage{algorithmic}
\usepackage{algorithm}
\usepackage{array}
\usepackage[caption=false,font=normalsize,labelfont=sf,textfont=sf]{subfig}
\usepackage{textcomp}
\usepackage{stfloats}
\usepackage{url}
\usepackage{verbatim}
\usepackage{graphicx}
\usepackage{standalone} 
\usepackage{cite}
\usepackage{makecell}
\usepackage{tikz}
\usepackage{soul}
\usepackage{xcolor}
\sethlcolor{yellow}

\usepackage{nomencl}
\makenomenclature

\newcommand{\rt}[1]{\textcolor{black}{#1}}
\newcommand{\rrtt}[1]{\textcolor{black}{#1}}

\newcolumntype{L}{l!{\vrule width 2pt}}

\DeclareMathOperator{\sign}{sign}

\usepackage{pgfplots}
\usepgfplotslibrary{groupplots} 
\pgfplotsset{compat=1.18}
\usepgfplotslibrary{fillbetween}
\usetikzlibrary{backgrounds}
\usetikzlibrary{patterns}
\usetikzlibrary{arrows.meta}
\tikzset{arrow/.style={-Latex, line width=1pt}}

\usetikzlibrary{angles, quotes, calc}

\usetikzlibrary{decorations.markings}

\usepackage{amsmath,amsbsy} 
\usepackage{bm} 
\usepackage{upgreek}  

\begin{document}

\title{\rt{Modeling Fault Recovery and Transient Stability of Grid-Forming Converters Equipped With Current Reference Limitation}}

\author{Ali Arjomandi-Nezhad,~\IEEEmembership{Student Member,~IEEE,}
\and
Yifei Guo,~\IEEEmembership{Member,~IEEE,}
\and
Bikash C. Pal,~\IEEEmembership{Fellow,~IEEE,}
\and
Guangya Yang,~\IEEEmembership{Senior Member,~IEEE.}

\thanks{This work was supported in part by the European Union’s Horizon 2020 Research and Innovation Programme under the Marie Skłodowska-Curie Grant 956433 (InnoCyPES Project), in part by the Resilient Operation of Sustainable Energy Systems (ROSES) U.K.-China (EPSRC-NSFC) Programme on Sustainable Energy Supply under Grants EP/T021713/1 and NSFC-52061635102, and in part by the Royal Society under Grant RG\textbackslash R2\textbackslash 232398. For the purpose of open access, the authors have applied a Creative Commons Attribution (CC BY) license to any Accepted Manuscript version arising.
}
\thanks{A. Arjomandi-Nezhad (corresponding author) and B. C. Pal are with the Department of Electrical and Electronics Engineering at Imperial College London, London SW7 2AZ, United Kingdom (e-mails: a.arjomandi nezhad21@imperial.ac.uk; b.pal@imperial.ac.uk).
}
\thanks{Y. Guo is with the Key Laboratory of Power System Intelligent Dispatch and Control of Ministry of Education, Shandong University, Jinan 250061 (e-mail: yifei.guo@sdu.edu.cn).}

\thanks{ G. Yang is with the Department of Wind and Energy Systems at the Technical University of Denmark, Anker Engelunds Vej 1, Bygning 101A, 2800 Kongens Lyngby, Denmark (e-mail: gyyan@dtu.dk).}}

\markboth{Submitted to IEEE Transactions on Energy Conversion}%
{Shell \MakeLowercase{\textit{et al.}}: A Sample Article Using IEEEtran.cls for IEEE Journals}


\maketitle

\begin{abstract}
When grid-forming (GFM) inverter-based resources (IBRs) face severe grid disturbances (e.g., short-circuit faults), the current limitation mechanism may be triggered. Consequently, the GFM IBRs enter the current-saturation mode, inducing nonlinear dynamical behaviors and posing great challenges to the post-disturbance transient angle stability. This paper presents a systematic study to reveal the fault recovery behaviors of a GFM IBR and identify the risk of instability. A closed-form expression for the necessary condition that a GFM IBR returns from the current-saturation mode to the normal operation mode is presented. Based on these analyses, it is inferred that the angle of the magnitude-saturated current significantly affects the post-fault recovery and transient stability; with different angle selection, the system may follow multiple post-fault trajectories depending on those conditions: 1) Convergence to a normal stable equilibrium point (SEP), 2) convergence to a saturated stable equilibrium point (satSEP), or 3) divergence (instability). In this paper, the circumstances under which a GFM IBR cannot escape from the current-saturation mode are thoroughly investigated. The theoretical analyses are verified by dynamic simulations.
\end{abstract}

\begin{IEEEkeywords}
Current limitation, current saturation, grid-forming (GFM) converters, post-fault recovery, transient stability, virtual synchronous generator (VSG).
\end{IEEEkeywords}

\nomenclature{APC}{Active power controller}
\nomenclature{RPC}{Reactive power controller}
\nomenclature{SEP}{Stable equilibrium point}
\nomenclature{satSEP}{Saturated stable equilibrium point}
\nomenclature{GFM}{Grid-forming}
\nomenclature{IBR}{Inverter-based resource}
\nomenclature{CRS}{Current reference saturation}
\nomenclature{CACRS}{Constant angle current reference saturation}
\nomenclature{DOA}{Domain of attraction}
\nomenclature{VSG}{Virtual synchronous generator}
\nomenclature{VI}{Virtual impedance}

\nomenclature{$\alpha$}{$90^\circ-\phi$}
\nomenclature{$\beta$}{Angle of the saturated current}

\nomenclature{$\phi$}{Phase angle of the total impedance from the connection point}
\nomenclature{$Z$}{Total impedance from the connection point}
\nomenclature{$R$}{Total resistance from the connection point}
\nomenclature{$X$}{Total reactance from the connection point}
\nomenclature{$L_f$}{The filter inductance}
\nomenclature{$C$}{The filter capacitance}
\nomenclature{$y$}{Shunt admittance of the filter}
\nomenclature{$V_d^{\rm sat}$}{Voltage of connection point along d-axis in the current-saturation mode}
\nomenclature{$V_q^{\rm sat}$}{Voltage of connection point along q-axis in the current-saturation mode}
\nomenclature{$V_d$}{Projection of the voltage of connection point on d-axis}
\nomenclature{$V_q$}{Projection of the voltage of connection point on q-axis}

\nomenclature{$V_g$}{Thevenin impedance of the grid}
\nomenclature{$V_d^{\rm ref}$}{Voltage reference in d-direction}
\nomenclature{$V_q^{\rm ref}$}{Voltage reference in q-direction}

\nomenclature{$i_{sd}$}{Inverter's current along d-axis}
\nomenclature{$i_{sq}$}{Inverter's current along q-axis}

\nomenclature{$i_{d}$}{Grid-side current along d-axis}
\nomenclature{$i_{q}$}{Grid-side current along q-axis}

\nomenclature{$i_{sd}^{\rm ref}$}{Current reference in d-direction}
\nomenclature{$i_{sq}^{\rm ref}$}{Current reference in q-direction}

\nomenclature{$\overline{i_{sd}^{\rm ref}}$}{Saturated current reference in d-direction}
\nomenclature{$\overline{i_{sq}^{\rm ref}}$}{Saturated Current reference in q-direction}

\nomenclature{$I_{s}^{\rm max}$}{The magnitude of the saturated current}

\nomenclature{$\delta$}{Active power controller (APC) angle ($\theta(t)-\theta_g(t)$)}
\nomenclature{$\delta_{af}$}{Post-fault angle}
\nomenclature{$\delta_{0}$}{Pre-fault angle}
\nomenclature{$\delta_d^p$}{The maximum APC angle such that $V_d^{\rm ref} \le V_d$}
\nomenclature{$\delta_{\rm sat}$}{The APC angle after which the GFM IBR enters the current saturation mode}

\nomenclature{$\delta_q^p$}{The minimum APC angle such that $V_q \le 0$}
\nomenclature{$\delta^{SE}$}{The angle of stable equilibrium point (SEP)}
\nomenclature{$\delta_s^{SE}$}{The angle of satSEP}
\nomenclature{$\delta_1^{UE}$}{The angle of unstable equilibrium 1}
\nomenclature{$\delta_2^{UE}$}{The angle of unstable equilibrium 2}

\nomenclature{$\mathcal{S}$}{The set of entering angles}
\nomenclature{$\mathcal{R}$}{The set of returning angles}

\nomenclature{$u_d$}{Output of the voltage PI controller in d-direction}
\nomenclature{$u_q$}{Output of the voltage PI controller in q-direction}
\nomenclature{$u_{\rm max}$}{Maximum value of the output of the voltage PI controller}

\nomenclature{$P$}{Active power output}
\nomenclature{$P_0$}{Active power reference}
\nomenclature{$D_p$}{Active droop coefficient}
\nomenclature{$H$}{Virtual inertia}
\nomenclature{$\omega_0$}{Set-point frequency}
\nomenclature{$\omega_n$}{Nominal frequency}
\nomenclature{$\omega$}{Frequency generated by APC}

\rrtt{\printnomenclature}

\section{Introduction}
\IEEEPARstart{R}{etirement} of conventional synchronous generator-based power plants in favor of inverter-based resources (IBRs) made it necessary to implement grid-forming (GFM) control to provide various services, e.g. frequency and voltage support \cite{Zhuang2022, Zhang2021, Ravanji_intro, Arjomandi2024}. Nevertheless, GFM IBRs equipped with the current limitation/saturation exhibit complicated nonlinear dynamics during and after large disturbances \cite{Guangya2021}, affecting the system stability. Therefore, it is vital to study and model their responses to large disturbances.

The ability of GFM IBRs to retain synchronism with the grid after being subjected to a large disturbance is defined as their transient stability \cite{Luo2022}. Since a GFM IBR usually synchronizes itself to the grid through the active power controller (APC), its transient stability can be redefined as the stability of the \rt{angle and frequency of} APC around the stable equilibrium point (SEP) \cite{Arjomandi2024},\rt{\cite{Gu2022}}. Several papers studied the transient stability of GFM IBRs around the SEP by adopting Lyapunov-based methods \cite{Lei2023_1, Fu2020}, Krylov–Bogoliubov–Mitropolsky asymptotic method \cite{Lei2023_2}, quantitative parameters constraining \cite{He2023, Liu2022}, and damping energy visualization and geometry approximation \cite{Lei2023_3}. However, these papers neglected the impact of current saturation on the dynamics of GFM IBRs. \rt{Unlike SGs, GFM IBRs usually have rather limited overcurrent capability, which is between 1.1 and 1.4 per unit (p.u.) if the GFM IBRs are not oversized \cite{Rajaei2023}, and therefore very easy to hit during disturbances.} It has been demonstrated in the literature that the current saturation adversely affects the transient stability \cite{Guangya2021, Arjomandi2024}.

\rt{Relatively fewer papers considered the current limitation} in transient stability analysis. Generally, there are two categories of current limitation methods \cite{Fan2022_1}: virtual impedance (VI) and current reference saturation (CRS). The former reduces the GFM voltage reference, whereas the latter directly limits the current reference generated by the voltage controller. \rt{References \cite{Khan2022, Qoria2020_1, Xiong2021, Zhang2023} studied transient stability of the VI-based current-limited GFM IBRs.} Although the VI-based methods preserve the voltage source behavior, they may fail to limit the current within the first few milliseconds \cite{Fan2022_1, Qoria2020_2}. In contrast, CRS-based methods limit the current promptly. Activation of CRS causes GFM IBRs to act as a current source while still being synchronized to the grid through the active power controller. 

\rt{CRS can be implemented in several ways. Circular CRS, which generates a reference with maximum magnitude and the angle of unsaturated current, is usually used in cascade with a virtual admittance voltage controller \cite{Fan2022_3}. Circular CRS is not usually cascaded with other types of voltage controllers. Other common types of CRS methods are d-axis priority CRS  \cite{Lyu}, q-axis priority CRS  \cite{Quedan2023}, and constant angle CRS (CACRS)\footnote{Constant angle in this context means a fixed angle from the local reference frame. This angle might vary from the common reference frame.} \cite{Qoria2020_3}. \rrtt{So far, a consensus has not been reached} on the most favorable current saturation strategies. The analysis of this paper applies to GFM IBRs equipped with CACRS. The transient stability of GFM IBRs with other CRS methods can be referred to \cite{Guangya2021, Fan2022_3, Huang2017, Wang2023, Ravanji2023}.}

\rt{Reference \cite{Lu2024} presented a segmental equal area criterion for transient stability assessment of virtual synchronous generator (VSG) GFM IBRs equipped with CACRS. A Lyapunov-based method for transient stability analysis of GFM IBRs equipped with CACRS is provided in \cite{Li2023}. The existence of a saturated stable equilibrium point (satSEP) is pointed out in \cite{Lu2024, Li2023, Rokrok2022}.} According to \cite{Rokrok2022}, if the angle of \rt{satSEP} is more than the saturation threshold angle, the GFM IBR might be locked into the current-saturation mode through converging to satSEP. Nevertheless, as will be shown later in this paper, there exists another circumstance for converging to the \rt{satSEP}. To explore circumstances in which a GFM IBR is locked into \rt{the satSEP, the conditions for entering and exiting} the current-saturation mode should be identified. Reference \cite{Fan2022_2} discussed that these two conditions are functions of the angle of the converter's reference dq-frame \rt{(local reference frame)} from the grid's Thevenin voltage angle \rt{(common reference frame)}. \rt{Reference \cite{Fan2022_2} also discussed that the set of angles for which the GFM IBR enters the current-saturation mode and the set of angles it exits are not complements of each other. In the rest of this paper, these two sets are mentioned as} (\emph{a}) the set of returning angles as the set of angles in which the voltage controller generates a current reference less than the allowable amount if it is in the current-saturated operation mode and (\emph{b}) the set of entering angles as the set of angles in which the GFM IBR transits from normal operation mode\rt{, where the current limitation mechanism is not activated,} to the current-saturation mode. Formulations for the set of entering angles have been extensively presented in the literature, e.g. \rt{in} \cite{Arjomandi2024},\rt{\cite{Lu2024}}. However, an analytical approach for the set of returning angles has not been attempted.

To the best of our knowledge, a closed-form expression for the set of returning angles and a comprehensive analysis of the situations in which a GFM IBR is locked into the current-saturation mode after disturbances are missing in the literature. To bridge this gap, this paper first identifies conditions under which a GFM IBR returns from the current-saturation mode. All the circumstances under which a GFM IBR converges to the \rt{satSEP} are also explored in this paper. All formulations and analyses provided in this paper are valid for different $X/R$ ratios. The main contributions of this paper are:
\begin{itemize}
    \item  A formulation for the set of returning angles is derived. It is proven that this set is a function of the angle of the magnitude-saturated current (mentioned as the saturated current in the rest of the paper).  

    \item The circumstances in which a GFM IBR is locked into the \rt{satSEP} are investigated.

    \item The influence of various parameters, such as the saturated current angle and the $X/R$ ratio of the grid impedance on the post-fault recovery and the transient stability, is analyzed.

\end{itemize}

The rest of the paper is organized as follows. Section \ref{sec:GFM_Model} presents a general overview of multi-loop VSG GFM IBRs. \rt{Section \ref{sec:CS_OM} formulates the situations in which a GFM IBR returns from the current-saturation mode. The causes of entering the current-saturation mode and the model of a GFM IBR in the current-saturation mode are also discussed in Section \ref{sec:CS_OM}.} The model of transient stability of a VSG considering \rt{CACRS} is provided in Section \ref{sec:TS}. Circumstances that a GFM IBR is locked into the current-saturation mode are also explored in Section \ref{sec:TS}. A sensitivity analysis on the effect of the grid's conditions and controller's parameters on transient stability is provided in Section \ref{sec:EPT}. The case studies are presented in Section \ref{sec:CS} followed by conclusions. 

\begin{figure*}[!t]
    \centering
    \includegraphics[width=0.7\linewidth]{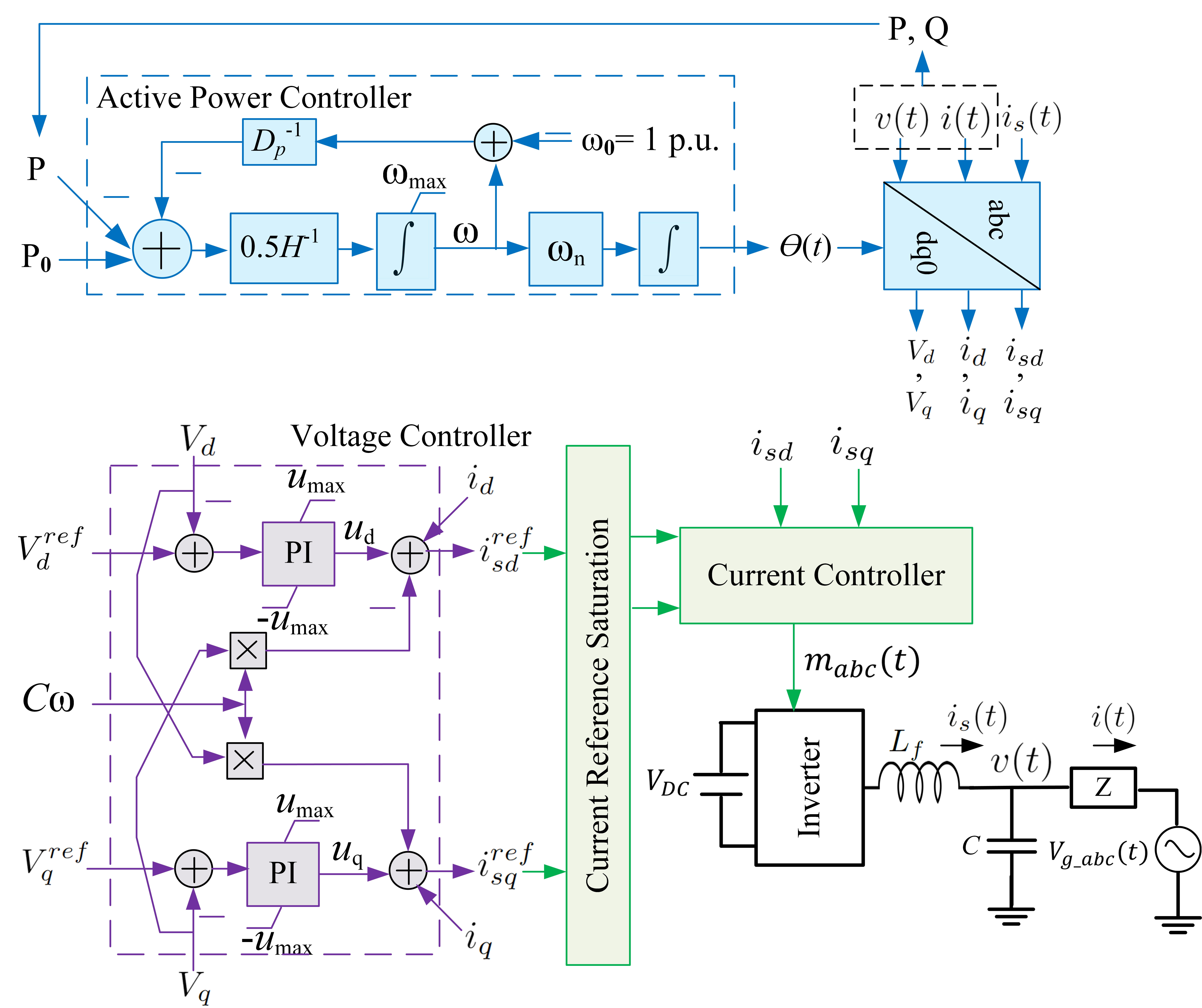}
    \caption{\rrtt{The structure of a three-layer VSG GFM IBR \cite{Arjomandi2024, Yazdani, Wang2023}.}}
    \label{fig:GridForming}
\end{figure*}

\section{System Structure}
\label{sec:GFM_Model}

There are several control structures for GFM IBRs in the literature. This paper considers a typical three-\rt{layer} VSG as shown in Fig.~\ref{fig:GridForming} \cite{Arjomandi2024, Yazdani},\rt{\cite{Wang2023}}. The inverter is connected to the grid, which is modeled as a Thevenin equivalent of voltage $\overrightarrow{V_g}$ and impedance of $R_g+jX_g$, through a transformer with the impedance $R_{tr}+jX_{tr}$ and an LC filter with inductance of $L_f$ and capacitance of $C$. The total equivalent impedance from IBR terminal is $(R_g+jX_g) + (R_{tr}+jX_{tr}) = R + jX = Z e^{j\phi}$. Since the filter capacitance is small, its effect is neglected in the rest of the paper. 

The inverter is controlled through a three-\rt{layer} hierarchical control containing the outer layer, the inner layer, and the innermost layer shown with colors blue, purple, and green in Fig.~\ref{fig:GridForming}, respectively. The outer layer consists of APC, which emulates the second order swing equation \cite{Kundur}, and reactive power controller (RPC), which controls the reactive power by setting a reference magnitude for the voltage \cite{Zhang2021}. Since RPC is not the focus of this paper, it is not shown in Fig.~\ref{fig:GridForming}, and we assume that it always generates the nominal voltage reference. APC generates the frequency and angle for the converter's reference dq-frame. 
 The difference between the angle of the converter's (local) reference frame and the common reference frame ($\delta := {\theta}-{\theta}_g$) is called APC angle, which plays a critical role in keeping the GFM IBR synchronous with the grid. \rrtt{The inputs to APC are $P_0$ and $P$, which are, respectively, the active power reference and active power output. Active power output is calculated from measured voltage and current output.} The inner controller includes the voltage controller which controls the terminal voltage $\overrightarrow{V}$ to be aligned with the d-axis with the magnitude of $V_d^{\rm ref}$. This controller generates the current reference for the current controller, which is the innermost control layer. The GFM IBR acts as a voltage source unless the magnitude of the current reference exceeds the \rt{threshold} $I_s^{\rm max}$. In this situation, the saturation block gives the saturated current reference to the current controller, and the GFM IBR behaves as a current source.

\section{Current-Saturation Operation Mode}
\label{sec:CS_OM}

\rt{In this section, the causes of entering the current-saturation mode and the corresponding modeling of a GFM IBR in the current-saturation mode are detailed. Then, a mathematical expression for the set of APC angles in which the GFM IBR returns to the normal operation mode is derived. This is followed by analyzing the operation mode of GFM IBRs for different APC angles.}

\begin{figure}[t]
    \centering
    \includegraphics[width=\linewidth]{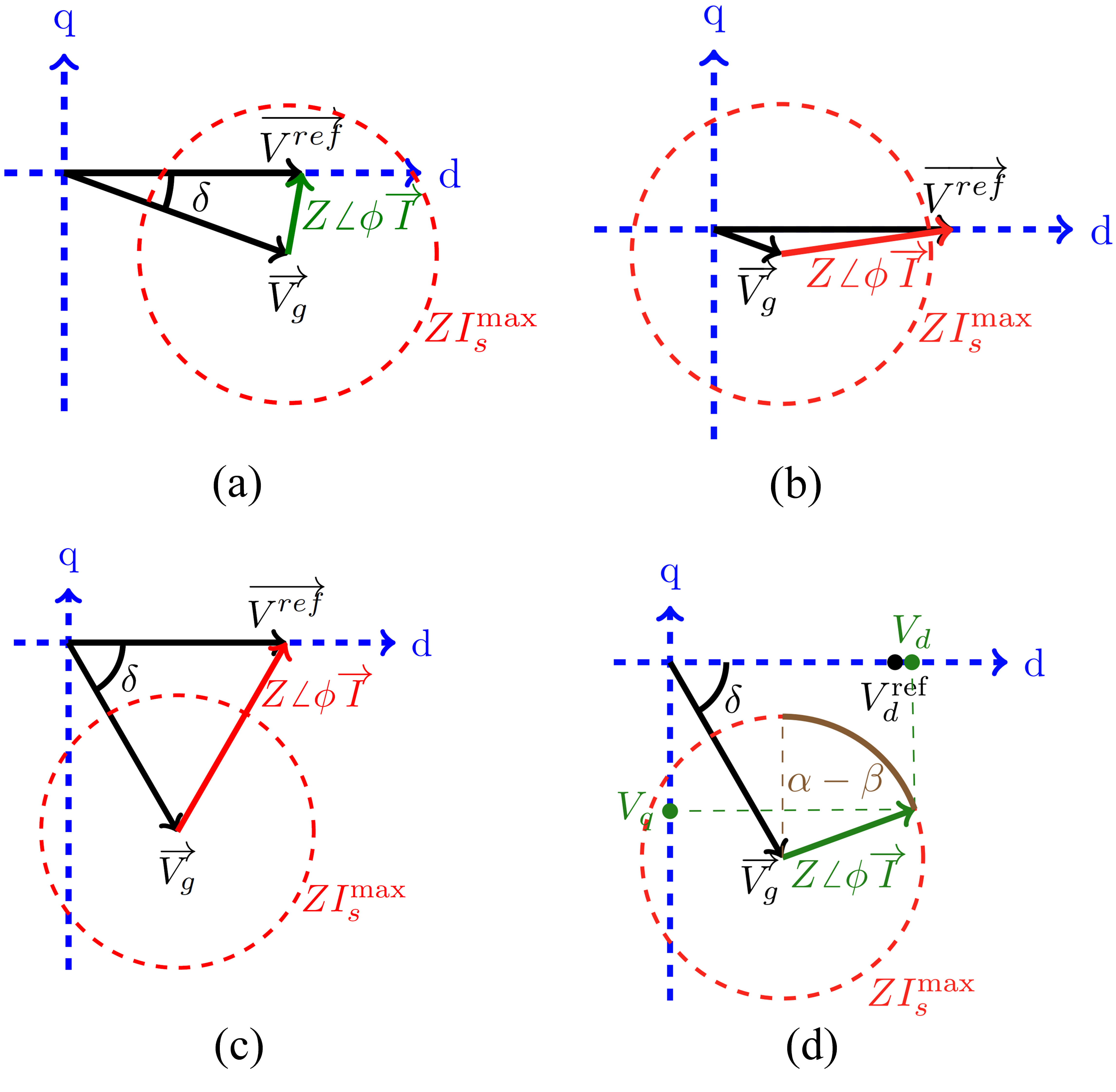}
    \caption{Phasor-Diagram of the voltages in (a) steady-state normal operation mode, (b) voltage sag, (c) extra APC angle, and (d) extra APC angle with CRS.}
    \label{fig:DQSAT}
\end{figure}

When a GFM IBR operates in the normal operation mode, its terminal voltage aligns with the d-axis and has the magnitude of $V_d^{\rm ref}$ as shown in Fig.~\ref{fig:DQSAT}(a). In this situation, the active power output is expressed as
\begin{align} 
\label{eq:Punsat}
    P_{\rm unsat} = \cfrac{(V_d^{\rm ref})^2}{Z}\sin{\alpha} + \cfrac{V_gV_d^{\rm ref}}{Z}\sin{(\delta - \alpha)}
\end{align} 
where $\delta$ = \({\theta}\)$(t)-$\({\theta}\)$_g(t)$ is the APC angle, which is the angle between the converter's dq-frame and grid's Thevinen voltage, and \rt{$\alpha = \arctan({R/X}) = 90^\circ-\phi$}. If there is a deep voltage sag Fig.~\ref{fig:DQSAT}(b) or excessive APC angle Fig.~\ref{fig:DQSAT}(c), the GFM IBR is no longer able to regulate the terminal voltage to \rt{$\overrightarrow{V_{\rm ref}} = (V_d^{\rm ref}, 0)$} subject to the current limit \cite{Arjomandi2024},\rt{\cite{Li2023}}. \rt{Therefore, a current limitation mechanism is needed to limit the current. In this paper, the studied current limitation mechanism is CACRS, in which the angle of saturated current is constant from the local (converter's) reference frame. When the magnitude of the current reference generated by the voltage controller surpasses the limitation, the saturation block gives the saturated current reference $ 
 (\overline{i_{sd}^{\rm ref}} , \overline{i_{sq}^{\rm ref}} ) = (I_s^{\rm max}\cos\beta, I_s^{\rm max}\sin\beta)$ to the current controller, where $\beta$ is the saturated current angle with respect to the converter's reference d-axis \cite{Rokrok2022}.} In this situation, $(V_d, V_q) \neq (V_d^{\rm ref}, 0)$ as depicted in Fig.~\ref{fig:DQSAT} (d). \rt{Neglecting the impact of the filter capacitor, according to kirchhoff's  voltage law the terminal voltage in the current-saturation mode in the local reference frame is
 \begin{align}
   \label{eq:kirchhoff}
   V_d^{\rm sat}+jV_q^{\rm sat}&=V_ge^{-j\delta}+I_s^{\rm max}e^{j\beta}Ze^{j(\cfrac{\pi}{2}-\alpha)}.
 \end{align}
After separating the real and imaginary parts of \eqref{eq:kirchhoff}, the terminal voltage of the IBR under the converter's reference dq-frame is calculated as}
\begin{align} 
\label{eq:Vdsat}
    V_d^{\rm sat} &= V_g\cos{\delta} + ZI_s^{\rm max}\sin(\alpha-\beta)\\
\label{eq:Vqsat}
    V_q^{\rm sat} &= -V_g\sin{\delta} + ZI_s^{\rm max}\cos(\alpha-\beta).
\end{align} 
Since the terminal voltage of the GFM IBR is no longer aligned with the d-axis of the local \rt{reference frame} in the current-saturation mode, the current angle $\beta$  no longer naturally reflects \rt{the} power factor. Instead, the power factor angle is $\arctan({V_q^{\rm sat}}/{V_d^{\rm sat}})-\beta$.

The active power output during the current-saturation operation mode is 
\begin{align} 
\label{eq:Psat}
    P_{\rm sat} &= R(I_s^{\rm max})^2+V_gI_s^{\rm max}\cos(\delta+\beta)
\end{align} 
which depends on both $\delta$ and $\beta$. Hence, the \rt{selection of $\beta$ significantly impacts the} transient behavior and therefore the stability of a GFM IBR, which will be detailed later in Sections \ref{sec:TS} and \ref{sec:EPT}.

\subsection{Sets of Entering and Returning Angles} \label{subsec:SEA}
The GFM IBR enters the current-saturation mode \cite{Arjomandi2024}, \rt{\cite{Lu2024, Wang2023}} if 
\begin{align} 
\label{eq:Theta_Sat}
   \cos\delta \leq \cfrac{1}{2}\left(\cfrac{V_d^{\rm ref}}{V_g}+\cfrac{V_g}{V_d^{\rm ref}} -\cfrac{(ZI_s^{\rm max})^2}{V_gV_d^{\rm ref}}\right).
\end{align} 
Defining the saturation threshold angle as
\begin{align} 
\label{eq:Theta_Sat_est}
    \delta_{\rm sat}: =\arccos\left(\cfrac{1}{2}\left(\cfrac{V_d^{\rm ref}}{V_g}+\cfrac{V_g}{V_d^{\rm ref}}-\cfrac{(ZI_s^{\rm max})^2}{V_gV_d^{\rm ref}}\right)  \right),
\end{align} 
the set of entering angles (denoted $\mathcal{S}$) where the APC angle satisfying the condition \eqref{eq:Theta_Sat}, is given by
\begin{align} 
\label{eq:Enter_set}
\mathcal{S} = [-180^\circ, \ -\delta_{\rm sat}] \cup [\delta_{\rm sat}, \ 180^\circ].
\end{align} 
This means a GFM IBR enters the current-saturation mode if the absolute value of the APC angle $\delta$ exceeds a threshold. This threshold depends on the grid voltage as analyzed in \cite{Arjomandi2024}, \rt{\cite{Lu2024}} in detail. 

As soon as the APC angle exits $\mathcal{S}$, the voltage controller of GFM IBR should generate a current reference less than the maximum allowable value, and the GFM IBR should come back to the normal operation mode. However, this may not be true under certain circumstances. The set of APC angles at which the GFM IBR can return to the normal operation mode spontaneously is expressed as
\begin{align} 
\label{eq:Rertun_set_def}
\mathcal{R} := \left\{ -180^\circ \le \delta \le 180^\circ \mid \left(i_{sd}^{\mathrm{ref}}\right)^2 + \left(i_{sq}^{\mathrm{ref}}\right)^2 \le (I_s^{\mathrm{max}})^2 \right\}
\end{align} 
where $i_{sd}^{\mathrm{ref}}$ and $i_{sq}^{\mathrm{ref}}$ are functions of the terminal voltage, which is, in turn, a function of $\delta$. These two references are unsaturated immediate outputs of the voltage controller.

One of the main goals of this work is to delve into $\mathcal{R}(\beta)$ by solving \eqref{eq:Rertun_set_def}. The current reference \rt{generated by the voltage controller} is given by \cite{Yazdani}:
\begin{align} 
\label{eq:Reference_Current_D}
    i_{sd}^{\rm ref} &= u_d-yV_{q} + i_d  
    \approx u_d + I_s^{\rm max}\cos{\beta} 
\end{align} 
\begin{align} 
\label{eq:Reference_Current_Q}
    i_{sq}^{\rm ref} &= u_q +yV_{d} + i_q 
    \approx u_q + I_s^{\rm max}\sin{\beta}
\end{align} 
where $y=C\omega$ is the filter shunt admittance, and $u_d$ and $u_q$ are the outputs of the PI voltage controller depicted in Fig.~\ref{fig:GridForming}. \rt{In the normal operation mode, the output of the PI controller are $u_d=C \cfrac{\text{d}V_d}{\text{d}t}$ and $u_q=C \cfrac{\text{d}V_q}{\text{d}t}$ \cite{Yazdani}. Therefore, these two variables are zero in the steady-state of the normal operation mode.} By substituting \eqref{eq:Reference_Current_D} and \eqref{eq:Reference_Current_Q}, into \eqref{eq:Rertun_set_def}, it can be rewritten as
\begin{align} 
\label{eq:condition_for_r}
    (u_d+ I_s^{\rm max}\cos{\beta})^2 + (u_q + I_s^{\rm max}\sin{\beta})^2 \le (I_s^{\rm max})^2.    
\end{align} 
\rt{This is equivalent to
\begin{align}
      2u_dI_s^{\rm max} \cos{\beta} + 2u_qI_s^{\rm max} \sin{\beta} + u_d^2 + u_q^2 \le 0.
     \label{eq:expanded_form}
\end{align}}
\rt{The output of the PI controllers of the voltage controller is limited between $-u_{\rm max}$ and $u_{\rm max}$ \cite{Wang2023, Huang2017}. Inside the PI controller block the clamping anti-windup method is employed to prevent the integrator from windup when the output of the block reaches its limits \cite{Wang2023, Huang2017, Huang2017_2}.
In the current-saturation mode, voltage error signals $V_d^{\rm ref}-V_d$ and $V_q^{\rm ref}-V_q$ are non-zero because the voltage is not regulated. Since the PI controllers are fast, they quickly reach their limitations upon the exposure to a sustained positive or negative error signal. Therefore,} $u_d$ and $u_q$ are either $u_{\rm max}$ or $-u_{\rm max}$ depending on the sign of the voltage error signals $V_d^{\rm ref}-V_d$ and $V_q^{\rm ref}-V_q$ where 
$V_q^{\rm ref}=0$. Therefore, equation \eqref{eq:expanded_form}, which is the condition for returning to the normal operation mode, can be rewritten as
\rt{\begin{align}
     \label{eq:condition_for_r_middle}
      2u_{\rm max}\sign(u_d)I_s^{\rm max} \cos{\beta} + 2u_{\rm max}\sign(u_q)I_s^{\rm max} \sin{\beta}\\ \nonumber
      \le -2u_{\rm max}^2,
\end{align}
which is equivalent to}
\begin{align} 
\label{eq:condition_for_r_equiv}
       \sign(u_d)\cos{\beta} + \sign(u_q)\sin{\beta} \le -\frac{u_{\max}}{I_s^{\rm max}}.
\end{align} 
$u_{\max}$ should be small \rt{in order to prevent a large rate of change of $V_d$ and $V_q$, which cause a huge transient in the capacitor's current.} If $u_{\max} \ll I_s^{\rm max}$, the right-hand side of \eqref{eq:condition_for_r_equiv} is close to zero. The set of the saturated current angle $\beta$, which satisfies \eqref{eq:condition_for_r_equiv}, is presented in Table \ref{Signs}.

\begin{table}[!b]
\begin{center}
\caption{The sets of $\beta$ satisfying \eqref{eq:condition_for_r_equiv}}
\label{Signs}
\begin{tabular}{ l|l|l}
\hline
\hline
 & $\sign(u_d)>0$  & $\sign(u_d)<0$  \\
\hline
$\sign(u_q)>0$  & $-90^\circ \le \beta \le -45^\circ$ & $-90^\circ \le \beta \le 0^\circ$  \\
\hline
 $\sign(u_q)<0$   &  $\emptyset$ & $-45^\circ \le \beta \le 0^\circ$ \\
\hline
\hline
\end{tabular}
\end{center}
\end{table}

According to Table \ref{Signs}, for $-45^\circ \le \beta \le 0^\circ$, \eqref{eq:condition_for_r_equiv} is satisfied if  $u_d$ is negative. Signals $u_d$ and $u_q$ have the same signs with the voltage controller's error signals, i.e., $V_d^{\rm ref}-V_d$ and $-V_q$, respectively. \rt{According to \eqref{eq:Vdsat}, $V_d^{\rm ref}-V_d$ is negative if $-\delta_d^p(\beta) \le \delta \le \delta_d^p(\beta)$, where $\delta_d^p(\beta)$ is defined as}
\begin{align} 
\label{eq:deltaDP}
    \delta_d^p(\beta) =\arccos\left(\cfrac{V_d^{\rm ref}-ZI_s^{\rm max}\sin{(\alpha-\beta)}}{V_g}\right).
\end{align} 
\rt{$\delta_d^p(\beta)$ is the maximum APC angle such that $V_d^{\rm ref} \le V_d$.} On the other hand, for $-90^\circ \le \beta \le -45^\circ$, \eqref{eq:condition_for_r_equiv} is satisfied if $u_q$ is positive, which means $-V_q>0$. According to \eqref{eq:Vqsat}, this is equivalent to $\delta_q^p(\beta) \le \delta \le 180^\circ-\delta_q^p(\beta)$, where $\delta_q^p(\beta)$ is defined as
\begin{align} 
\label{eq:deltaQP}
    \delta_q^p(\beta) =\arcsin\left(\cfrac{ZI_s^{\rm max}\cos{(\alpha-\beta)}}{V_g}\right). 
\end{align} 
\rt{$\delta_q^p(\beta)$ is the minimum APC angle such that $V_q \le 0$.} Therefore, $\mathcal{R}$ depends on $\beta$, that is,
\begin{align} 
\label{eq:Rertun_set}
\mathcal{R}(\beta) = 
\begin{cases} 
[-\delta_d^p(\beta), \ \delta_d^p(\beta)], & \text{if}\,\,\beta \in [-45^\circ, 0^\circ] \\
[\delta_q^p(\beta), \ 180^\circ- \delta_q^p(\beta)], &\text{if}\,\,  \beta \in [-90^\circ, -45^\circ].
\end{cases}
\end{align} 

 \rt{To reveal the rationale behind \eqref{eq:Rertun_set}, a comparison between the current reference and maximum allowed current for different ranges of $\beta$ is conducted. When $\beta \in [-45^\circ, 0^\circ]$, the dominant component of the current reference is $i^{\rm ref}_{sd}$, which is  positive. When APC angle $\delta$ enters the interval $[-\delta_d^p(\beta),\delta_d^p(\beta)]$, the sign of the d-component of voltage error becomes negative, therefore $u_d$ becomes negative. As a result,} the magnitude of the current reference becomes smaller than the maximum value according to \eqref{eq:Reference_Current_D} and \eqref{eq:condition_for_r}. \rt{When $\beta \in [-90^\circ, -45^\circ]$, $i^{\rm ref}_{sq}$, which is negative, is the dominant component of the current reference. As soon as $\delta$ enters the interval $[\delta_q^p(\beta), \ 180^\circ- \delta_q^p(\beta)]$, the q-component of voltage error and, consequently, $u_q$ become positive. Therefore, the magnitude of the current reference becomes smaller than the maximum value according to \eqref{eq:Reference_Current_Q} and \eqref{eq:condition_for_r}.} 

\begin{figure}[!t]
    \centering
    \includegraphics[width=\linewidth]{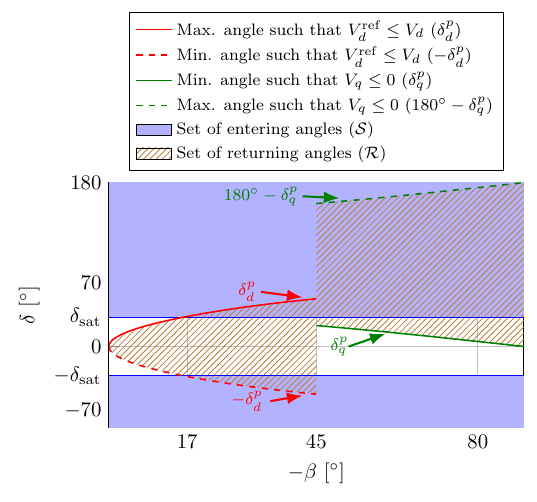}
    \caption{\rrtt{Illustration of entering and returning sets represented by the solid and hatched fill areas, respectively. Here, $Z=0.46$ p.u. and $X/R=\infty$.}}
    \label{fig:Entering_and_Returning}
\end{figure}
 
Fig.~\ref{fig:Entering_and_Returning} shows the variation of $\mathcal{R}$ and $\mathcal{S}$. $\mathcal{R}$ is a function of $\beta$ whereas $\mathcal{S}$ is independent from $\beta$. This is because while the GFM IBR is in the normal operation mode, the voltage error does not depend on the saturated current angle. Therefore, $\beta$ does not affect the current reference generated by the voltage controller in normal operation mode. Nevertheless, as soon as the GFM IBR enters the current-saturation mode, voltage error highly depends on the saturated current angle. Hence, $\beta$ significantly affects the magnitude of the current reference generated by the voltage controller. \rrtt{As depicted in Fig.~\ref{fig:Entering_and_Returning}, set of entering angles equals the set of angles in which $V_d^{\rm ref} \le V_d$ if $-45^\circ < \beta<0$, and it is the set of angles in which $V_q \le 0$ if $-90^\circ<\beta< -45^\circ$. This set is depicted with the hatched area in this figure. The filled area in Fig.~\ref{fig:Entering_and_Returning} visualizes the set of entering angles.}

\rt{ Depending on the APC angle, the GFM IBR might exhibit four different post-fault behaviors:}

\begin{itemize}
    \item $\delta \in \mathcal{S}-\mathcal{R}(\beta)$: The GFM IBR operates in the current-saturation mode \cite{Fan2022_2}. It is shown with the solid-filled area in Fig.~\ref{fig:Entering_and_Returning}. 
    \item $\delta \in \mathcal{R}(\beta)-\mathcal{S}$: The GFM IBR operates in the normal operation mode \cite{Fan2022_2}. The \rrtt{unfilled} hatched area in Fig.~\ref{fig:Entering_and_Returning} represents this subset. 
    \item $\delta \in \mathcal{R}(\beta) \cap \mathcal{S}$: The GFM IBR oscillates between normal operation and current-saturation modes \cite{Fan2022_2}. This subset is shown as the intersection of the solid-filled area and hatched area in Fig.~\ref{fig:Entering_and_Returning}. These oscillations can be reduced using the forced saturation introduced in \cite{Arjomandi2024}.
    \item $\delta \notin \mathcal{R}(\beta) \cup \mathcal{S}$: The GFM IBR retains its mode. The white area in Fig.~\ref{fig:Entering_and_Returning} represents this situation. If the GFM IBR was already in the current-saturation mode, it remains in the current-saturation mode even though the post-disturbance APC angle $\delta_{af} \notin \mathcal{S}$. 
\end{itemize}

\section{Transient Stability of GFM IBRs} \label{sec:TS}

The GFM IBR synchronizes itself to the grid through APC, which follows the second-order swing equation \cite{Kundur} if it is designed as a VSG:
 \begin{align}
    P_0-P
    &=
    2H \cfrac{d\omega}{dt}
    +
    \cfrac{1}{D_p} \left( \omega - \omega_0 \right) 
    \label{swing}
    \\
    \cfrac{d{\delta}}{dt}
    &=
    \omega_n \left( \omega - \omega_0 \right) \; 
    \label{deltaOmega}
\end{align}
where $H$, $D_p$, $P_0$, $\omega_0$, and $\omega_n$ are virtual inertia, active droop coefficient, active power reference, set-point frequency, and the nominal frequency, respectively. These equations outline the strong tie between the trajectory of the APC and the active power, which follows \eqref{eq:Punsat} and \eqref{eq:Psat} for the normal and the current-saturation operation modes, respectively. The intersections of the power-angle curve and the power reference, shown in Fig.~\ref{fig:Pposnegdelta}, form equilibrium points which are: 

\begin{figure}[t]
    \centering
        \includegraphics[width=\linewidth]{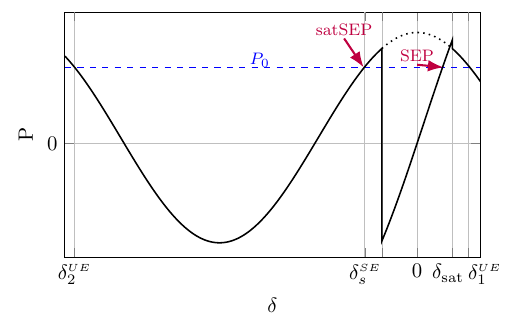}
   \caption{\rrtt{Power-Angle curve for a GFM IBR equipped with the CRS.}}
    \label{fig:Pposnegdelta}
\end{figure}

\begin{itemize}
    \item SEP: This is the normal stable equilibrium point to which convergence is desired. Deduced from \eqref{eq:Punsat}, it is calculated as
     \begin{align} 
\label{eq:SEP}
    \delta^{SE} =  \alpha+\arcsin\left(\cfrac{Z}{V_gV_d^{\rm ref}}\left(P_0-\cfrac{(V_d^{\rm ref})^2}{Z}\sin\alpha\right)\right).
\end{align} 

    \item Unstable Equilibrium 1 ($\delta_1^{UE}$): If the APC angle reaches this value while $\omega>\omega_0$, the GFM IBR loses its synchronism with the grid. Based on \eqref{eq:Psat}, it is calculated as 
 \begin{align} 
\label{eq:UE1}
    \delta_1^{UE}(\beta) &= - \beta + \arccos\left(\cfrac{P_0-R(I_s^{\rm max})^2}{V_gI_s^{\rm max}}\right).
\end{align} 
It shows that decreasing $\beta$ increases the $\delta_1^{UE}(\beta)$ which enhances transient stability by pushing away the unstable equilibrium point.  

    \item Unstable Equilibrium 2 ($\delta_2^{UE}(\beta)$): The GFM IBR loses its synchronism if the APC angle reaches this value while $\omega<\omega_0$. According to \eqref{eq:Psat}, this angle is calculated as
\begin{align} 
\label{eq:UE2}
    \delta_2^{UE}(\beta) = \delta_1^{UE}(\beta) - 360^\circ.
\end{align}
\rt{This equilibrium point is unlikely to be reached in practice except in some adverse situations, e.g. dramatic phase jump.}

    \item \rt{satSEP} ($\delta_s^{SE}(\beta)$): It is the stable intersection between the power reference and power output in the current-saturation mode. Deduced from \eqref{eq:Psat}, it is calculated as
    \begin{align} 
    \label{eq:satSEP}
        \delta^{SE}_s(\beta) &= - \beta - \arccos\left(\cfrac{P_0-R(I_s^{\rm max})^2}{V_gI_s^{\rm max}}\right).
    \end{align} 
    \end{itemize}

    It has been discussed in \cite{Rokrok2022} that if  $\delta_{\rm sat} \le \delta_s^{SE}(\beta)$, the \rt{satSEP} has a DOA that causes locking in the saturation mode. It will be discussed in the following subsection that there is another condition for locking into the saturation mode. \rt{Notice that since the power-angle curves are trigonometric functions, the equilibrium points repeat themselves every $360^\circ$. Hence, it is theoretically possible that a GFM IBR converges to another stable equilibrium point after passing an unstable equilibrium point. However, the power output becomes negative in a wide range of angles before it converges to a far stable equilibrium point. Such negative power might cause DC link capacitor over-voltage, which may trip the protection. Therefore, we only focus on analyzing the behavior of GFM IBR between the two closest unstable equilibrium points.}

\subsection{Converging to the Saturated Stable Equilibrium Point}\label{Subsec:CTSSEP}
\begin{figure}[b]
    \centering
        \includegraphics[width=\linewidth]{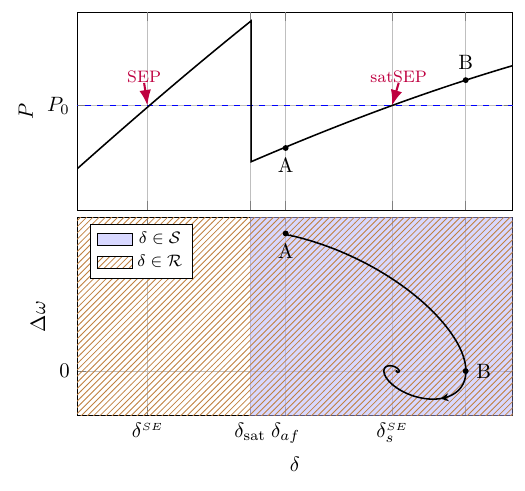}
  \caption{\rrtt{Post-fault trajectory and active power for the case where $\mathcal{C}_1$ holds.}}
    \label{fig:C1_exp_1}
\end{figure}
\rt{A GFM IBR is locked into the current-saturation mode if it converges to the satSEP. It has been shown in \cite{Rokrok2022} that if $\delta_s^{SE}(\beta) \in \mathcal{S}$ while the post-fault APC angle $\delta_{af} \in \mathcal{S}$, the post-fault trajectory might converge to the satSEP. However, such condition, which is denoted as $\mathcal{C}_1$ here, is not necessary for the convergence to the satSEP}. Since the set of returning angles $\mathcal{R}(\beta)$ is not exactly the complement of $\mathcal{S}$, there are cases that even though $\delta_s^{SE}(\beta) \notin \mathcal{S}$, the GFM IBR cannot escape from the current-saturation mode. \rt{It happens if the APC angle does not enter $\mathcal{R}(\beta)$ during its post-disturbance trajectory until converging to satSEP. Notice that in this condition it does not matter whether the APC angle trajectory is inside or out of $\mathcal{S}$. Since it does not enter $\mathcal{R}(\beta)$ it stays in the current saturation mode. This circumstance} is denoted as $\mathcal{C}_2$ here as the combination of the following conditions: (\emph{a}) $\delta_s^{SE}(\beta) \notin \mathcal{R}(\beta) \cup \mathcal{S}$, (\emph{b}) the post-fault trajectory toward $\delta_s^{SE}(\beta)$ does not enter the set of returning angles $\mathcal{R}(\beta)$, and (\emph{c}) the GFM IBR does not lose synchronism with the grid. This set of conditions is expressed as
\begin{align} 
\label{eq:C2Def}
\mathcal{C}_2: 
\begin{cases} 
\delta^{SE}_s(\beta) \notin \mathcal{R}(\beta) \cup \mathcal{S}  & (\mathcal{C}_{21})\\
\delta(t) \notin \mathcal{R}(\beta), \forall t &  (\mathcal{C}_{22}) \\
\delta_2^{UE}(\beta) < \delta(t) < \delta_1^{UE}(\beta), \forall t &   (\mathcal{C}_{23})
\end{cases}
\end{align} 
where $\mathcal{C}_{21}$ is an static condition, $\mathcal{C}_{22}$ and $\mathcal{C}_{23}$ are dynamic conditions.

\begin{figure}[!b]
    \centering
        \includegraphics[width=\linewidth]{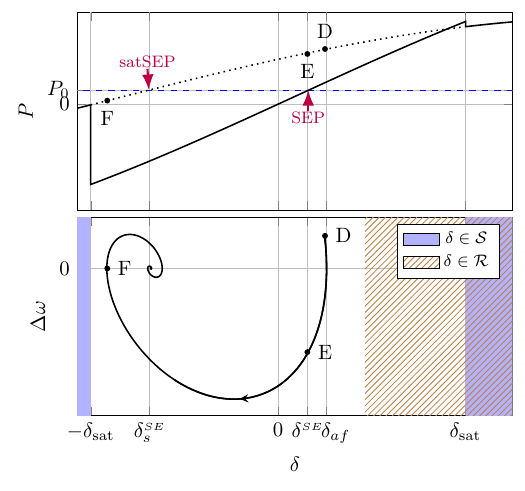}
  \caption{\rrtt{Post-fault trajectory and active power output for a scenario that condition $\mathcal{C}_2$ meets. The dotted curve is the power output in the current-saturation mode while $\delta \notin \mathcal{S}$.}}
    \label{fig:C2_exp_1}
\end{figure}

Figs.~\ref{fig:C1_exp_1} and~\ref{fig:C2_exp_1} depict two \rt{examples} of converging to the \rt{satSEP} for a GFM IBR with $D_p=0.03$ p.u. connected to a Grid with the equivalent impedance of $Z=0.46$ p.u. and $X/R=20$. In these figures, the solid-filled area represents the $\mathcal{S}$ and the hatched area represents $\mathcal{R}(\beta)$. The GFM IBR is forced to be in the current-saturation mode if $\delta \in \mathcal{S} \cap \mathcal{R}(\beta)$ as it was proposed in \cite{Arjomandi2024}. The convergence to \rt{the satSEP} in Fig.~\ref{fig:C1_exp_1} is due to $\mathcal{C}_1$. In this case, \rt{the satSEP} belongs to the set of entering to saturation angles and contains point A in its DOA. On the other hand, $\mathcal{C}_2$ \rt{is the cause of} locking into the saturation mode in Fig.~\ref{fig:C2_exp_1}. \rt{In this example,} \rt{the satSEP} is out of $\mathcal{S}$. However, since the post-fault angle ($\delta_D$) is out of $\mathcal{R}(\beta)$, the GFM IBR does not come back to the normal operation mode. \rt{The remainder of its post-fault trajectory does not pass any point in $\mathcal{R}(\beta)$ until it converges to the satSEP.} According to Fig.~\ref{fig:C2_exp_1}, the GFM IBR does not converge to the SEP even though the post-fault state D is close to SEP. The fulfillment of condition $\mathcal{C}_2$ imposes a significant challenge as even relatively small disturbances might also lead to undesired consequences.

\begin{figure}[!t]
    \centering
        \includegraphics[width=\linewidth]{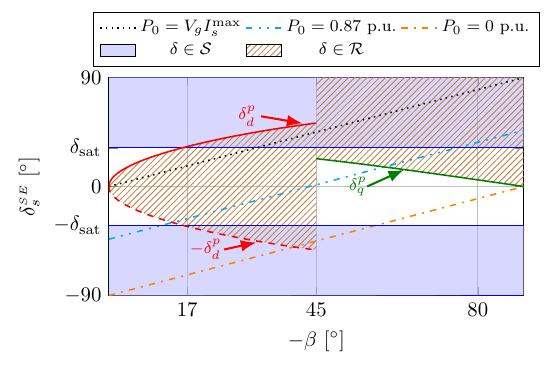}
  \caption{\rrtt{Relation between the satSEP angle and $\beta$. Solid filled area represents $\mathcal{S}$ and hatched area shows $\mathcal{R}(\beta)$.}}
    \label{fig:DeltaSEs}
\end{figure}

\rt{As discussed earlier in this section, risk of convergence to the satSEP highly depends on the value of $\delta^{SE}_s(\beta)$ and $\mathcal{R}(\beta)$, which} are functions of the saturated current angle $\beta$, the proper selection of $\beta$ can eliminate the risk of locking into the current-saturation mode through $\mathcal{C}_1$ or $\mathcal{C}_2$. \rt{If the sufficient condition $\delta^{SE}_s(\beta) \in \mathcal{R}(\beta) - \mathcal{S}$ is met}, the GFM IBR is not locked into the current-saturation mode. This is because of the fact that the GFM IBR comes back to the normal operation mode at the angle of \rt{satSEP}. $\delta^{SE}_s(\beta)$ depends on the power reference in addition to $\beta$ as demonstrated in Fig.~\ref{fig:DeltaSEs}. Therefore, $\beta$ satisfying the mentioned condition differs for different power references. Hence, an adaptive $\beta$ selection is needed to mitigate or decrease the risk of converging to the \rt{satSEP}.

To avoid the complexity in Section~\ref{sec:EPT}, it is assumed that $\mathcal{R}(\beta)$ and $\mathcal{S}$ are complements. Therefore, condition $\mathcal{C}_2$ is not analyzed in Section~\ref{sec:EPT}. This assumption is made only in that section. Later, cases that these two sets are not complements are analyzed in the Case Study section.

\section{Effect of Parameters on Transient Stability}\label{sec:EPT}
\begin{figure}[!t]
    \centering
        \includegraphics[width=\linewidth]{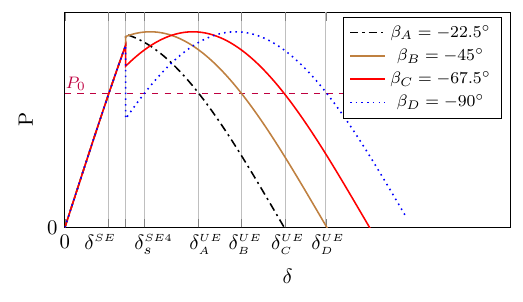}
  \caption{Power-Angle curves for different saturated current angles $\beta$.}
    \label{fig:PdiffBeta}
\end{figure}

It has been discussed in \cite{Arjomandi2024} that the current saturation deteriorates the post-fault \rt{transient stability} by decreasing the acceleration area and the angle of the \rt{unstable} equilibrium point 1. This decrement depends on $\beta$ in addition to APC control parameters and grid parameters because $\beta$ affects the \rt{active power-angle curve} as demonstrated in Fig.~\ref{fig:PdiffBeta}. This figure depicts that a smaller $\beta$ leads to a shift in unstable equilibrium point 1 and an increase in deceleration area compared to a bigger $\beta$. However, if $\beta$ exceeds a \rt{lower limit}, there exists an \rt{satSEP} in $\mathcal{S}$, which might cause locking in the current-saturation mode through the condition $\mathcal{C}_1$.  

\begin{figure*}[!t]
    \centering
    \includegraphics[width=0.7\linewidth]{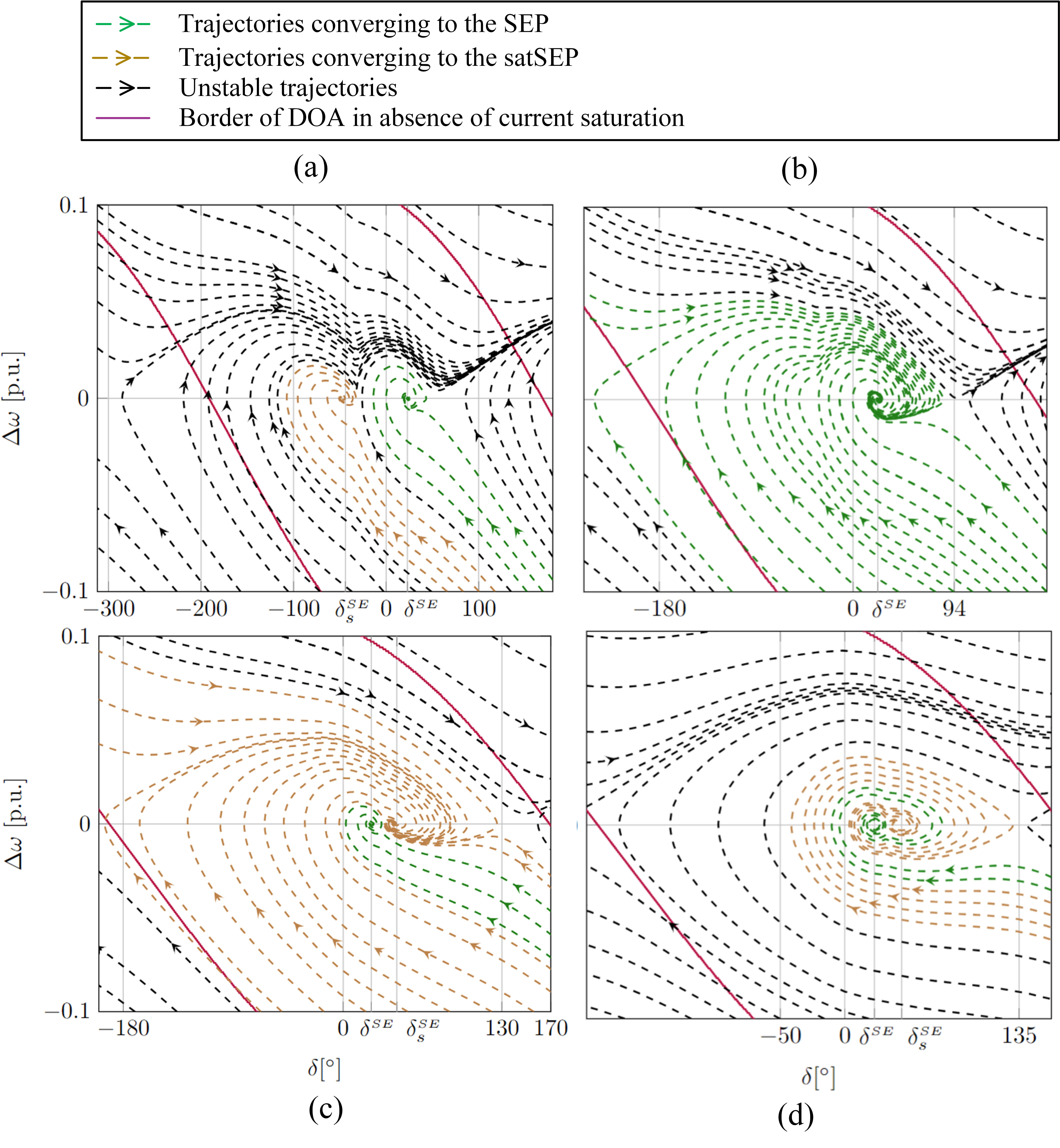}
    \caption{\rrtt{Phase portraits for a GFM IBR connected to an equivalent impedance of $Z=0.46$ p.u., and $X/R = 10$, and $\beta$ equals to (a) $0^\circ$, (b) $-45^\circ$, (c) $-90^\circ$, and (d) $-90^\circ$. $D_p$ equals to $0.03$ p.u. for (a)-(c) and $0.09$ p.u. for (d). Purple lines are borders of DOA for unsaturated GFM IBRs. Green lines and brown lines are trajectories converging to the SEP and the satSEP, respectively.}}
    \label{fig:PhasePortrateBeta}
\end{figure*}

Fig.~\ref{fig:PhasePortrateBeta} shows post-disturbance transient phase portraits for different saturated current angles. $\beta$ is zero in Fig.~\ref{fig:PhasePortrateBeta}(a). The \rt{satSEP} attracts a subset of the state-space under the condition $\mathcal{C}_1$. In Fig.~\ref{fig:PhasePortrateBeta}(b), $\beta$ is $-45^\circ$. Thus, \rt{the satSEP} is out of $\mathcal{S}$, and $\mathcal{C}_1$ cannot hold. The DOA of SEP is considerably bigger than that of Fig.~\ref{fig:PhasePortrateBeta}(a). The saturated current angle is  $-90^\circ$ in Fig.~\ref{fig:PhasePortrateBeta}(c). As depicted in this phase portrait, DOA of the \rt{satSEP} is considerably bigger than DOA of the SEP. If the droop coefficient increases from 0.03 p.u. to 0.09 p.u., the DOAs of these two equilibrium points become hardly separable as shown in Fig.~\ref{fig:PhasePortrateBeta}(d). Therefore, it is crucial to set $\beta$ in a way to exclude the \rt{satSEP} from $\mathcal{S}$. So that the condition $\mathcal{C}_1$ is not met.


\begin{figure}[!t]
    \centering
    \includegraphics[width=\linewidth]{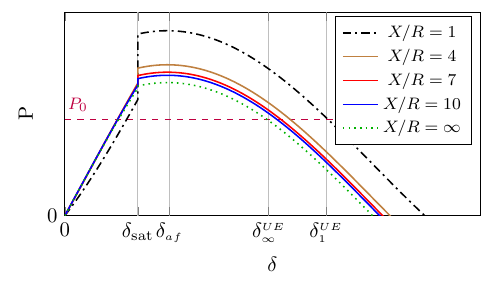}
   \caption{Power angle curves for different $X/R$ ratios. The GFM IBR is connected to an equivalent impedance of $0.46$ p.u., and $D_p=0.03$ p.u.}
    \label{fig:PdiffXperR}
\end{figure}

Other main effecting parameters are the droop coefficient, $X/R$ ratio, and the total impedance $Z$, which are analyzed in detail in the rest of this section. All DOAs are assessed through the numerical calculation of trajectories based on the swing equation and saturated and unsaturated active powers. $\beta$ in these cases is $-45^\circ$ so that condition $\mathcal{C}_1$ does not hold.


\begin{figure}[!t]
    \centering
  \includegraphics[width=\linewidth]{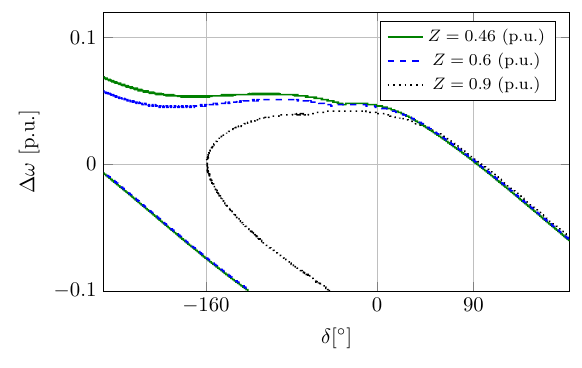}
   \caption{Borders of DOAs for different total impedances. \rt{$X/R=10$}, and $D_p=0.03$ p.u.}
\label{fig:SensevityToZ}
\end{figure}

\rt{Decreasing the droop coefficient provides more damping to the swing equation \eqref{swing}. Therefore, an expanded DOA is expected \cite{Lei2023_3}}. The total impedance $Z$ and $X/R$ ratio shape the relation between the power and angle. Therefore, they influence the dynamics of APC. Fig.~\ref{fig:PdiffXperR} depicts that the power output in the current-saturation mode is more for resistive grid impedance compared to inductive grids. \rt{This difference accounts for the active power losses.} Furthermore, the unstable equilibrium point 1 is shifted away. Therefore, the GFM IBR exhibit better transient stability when it is connected to a more resistive grid. The DOA of the SEP when the GFM IBR is connected to a weaker grid  (larger amount of impedance) is smaller compared to those connected to stronger grids according to Fig.~\ref{fig:SensevityToZ}.


\section{Case Studies} \label{sec:CS}

\begin{table}[!t]
\begin{center}
\caption{Parameters of the simulated GFM farm.}
\label{GFMbasedParameters}
\begin{tabular}{ llll}
\hline
\hline
\emph{Param.} & \emph{Values}  & \emph{Units} &  \emph{Descriptions} \\
\hline
$S_b$   & 310 & MVA & Nominal apparent power  \\
$V_b$ &  400    &  V &  Nominal Voltage\\
$V_{dc}$  &   1200   & V  &    DC link voltage   \\
n       & 816 & - & Number of \rt{identical} parallel GFM IBRs  \\
$f_n$   & 60 & Hz & Nominal frequency  \\
$D_p$   & 0.03    &  p.u. & Active droop coefficient  \\
$H$       & 2    &  p.u. &   Virtual inertia  \\
$\Delta\omega_{\rm max}$ & 0.0066     &  p.u. &   Maximum frequency deviation \\
$V_d^{\rm ref}$ & 1  &  p.u. & Voltage reference magnitude \\
$I_s^{\rm max}$ & 1.2     &  p.u. &   Maximum allowed current  \\
$X_{\rm tr}$ & 0.16     &  p.u. &   Total reactance of transformer windings \\
\rt{$u_{\rm max}$} & \rt{0.063}    & \rt{p.u.}  &   \rt{Max. output of the voltage PI controller} \\
\hline
\hline
\end{tabular}
\end{center}
\end{table}

To examine the analyses provided in the previous sections, the system of Fig.~\ref{fig:GridForming} with the general parameters of Table.~\ref{GFMbasedParameters} \cite{Arjomandi2024} is simulated in Simulink/MATLAB \rt{\cite{MATLAB}}. Since the GFM IBRs are parallel in the same bus, with the same sharing of loading, and identical, they can be modeled as an equivalent GFM IBR ~\cite{Tayyebi2020}. If the IBRs were not identical and in the same location, the analysis of this GFM farm would be more complicated. It is connected to a grid with the post-fault Thevenin impedance of $0.3$ p.u. and $X/R$ ratio of 20 (totally, $Z=0.46$ p.u.). Eight cases are simulated. In all cases, the system is assumed to be exposed to a three-phase short-circuit fault, which is modeled as a voltage drop of 0.95 p.u. until reaching the post-fault angle $\delta_{af}$ right after clearing the fault. These cases are \rrtt{listed in Table.~\ref{SimulatedCases}.}

\begin{table*}[!t]
\begin{center}
\caption{\rrtt{Simulated Cases.}}
\label{SimulatedCases}
\begin{tabular}{ llllllllll }
\hline
\hline
\emph{Case} & $P_0$ & \makecell{\emph{Fault duration} \\ \emph{[ms]}} & $\delta_0 [^\circ]$ &  $\delta_{af}[^\circ]$ & $\beta[^\circ]$  &   $\mathcal{R}[^\circ]$ &  $\delta^{SE}_s$  &  $\delta^{UE}_1[^\circ]$ \\
\hline
A & 270 MW (0.87 p.u.) & 100  & 23.38 & 34.93 & \textbf{-6} & \textbf{[-23.14, 23.14]} &  \textbf{-39.78} & \textbf{51.78}  \\
B & 270 MW& 100  & 23.38 & 34.93 & \textbf{-30} & \textbf{[-45.2, 45.2]} &  \textbf{-15.77} & \textbf{75.78}   \\
C &270 MW  & 100  & 23.38 & 34.93 & \textbf{-90} & \textbf{[-1.3,181.3]} &  \textbf{44.22} & \textbf{135.78}  \\
\hline
D &62 MW (0.2 p.u.)& \textbf{600}  & 5.23 & \textbf{44.76} & -60 & [14.84, 165.16] &  -22.00  & 142.00   \\
E &62 MW  & \textbf{100}  & 5.23 & \textbf{7.93} & -60 & [14.84, 165.16] &  -22.00 & 142.00   \\
\hline
F &270 MW (0.87 p.u.)  & \textbf{290} & 23.38 & \textbf{62.01} & -30 & [-45.2, 45.2] & -15.77 & 75.78   \\
G &270 MW & \textbf{330} & 23.38 & \textbf{67.71} & -30 &  [-45.2, 45.2] &  -15.77 & 75.78   \\
H &270 MW  & \textbf{400} & 23.38 & \textbf{76.10} & \multicolumn{4}{c}{\textbf{No current saturation implemented}}  \\

\hline
\hline
\end{tabular}
\end{center}
\end{table*}

Cases A, B, and C are compared to each other to demonstrate the effect of $\beta$ on post-fault recovery and transient stability. \rrtt{$\beta$ and the parameters that depend on $\beta$ are different for these three cases as mentioned in Table.~\ref{SimulatedCases}.} Cases D and E are compared to each other to investigate the risk of locking in the current-saturation mode \rrtt{through $\mathcal{C}_2$, explained in Subsection.~\ref{Subsec:CTSSEP}} when the post-fault angle is out of $\mathcal{R} \cup \mathcal{S}$. \rrtt{$\beta$ is the same for these two cases; however, fault duration and $\delta_{af}$ are different.} Case D is exposed to a longer duration of fault to reach the bigger post-fault angle $\delta_{af}$ right after clearing the fault. Cases F, G, and H are compared to show a reduction in DOA caused by the current saturation. $\beta$ and all other parameters are identical for Cases F and G. \rrtt{However, the fault duration for case G is longer.} The post-fault angle $\delta_{af}$ for Cases F and G are inside and outside the DOA, respectively. \rrtt{Fault duration} in Case H is larger than F and G, and the GFM IBR's current is not limited. \rrtt{Case H is simulated to demonstrate the transient stability of the GFM IBR in the absence of current saturation.} Since the grid impedance is the same for all cases, the set of entering angles, which is $\mathcal{S} = [-180^\circ, -32.0455^\circ] \cup [32.0455^\circ, 180^\circ]$, is identical for all of them except Case H, which is unsaturated. \rrtt{The parameters that get different values in each group of compared cases are in bold. For all these cases, fault started at $t_f=0.05$ s.} Detailed analyses of these cases are provided in the rest of this section.

\begin{figure}[!t]
    \centering
  \includegraphics[width=\linewidth]{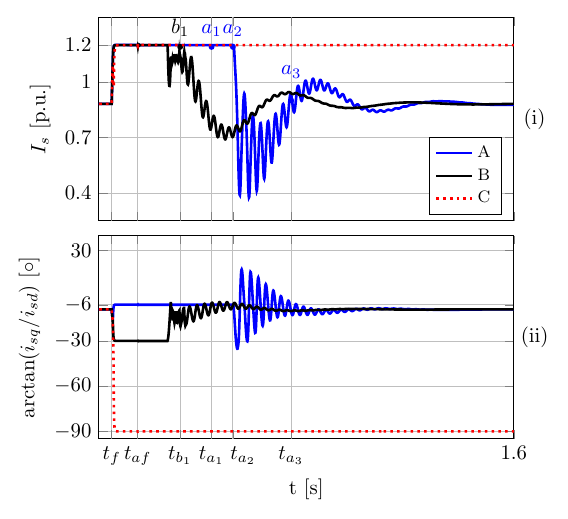}
   \caption{\rrtt{Simulation results for the post-fault currents in Cases A, B, and C.}}
    \label{fig:CurrentCaseABC}
\end{figure}

\begin{figure}[!t]
    \centering
  \includegraphics[width=\linewidth]{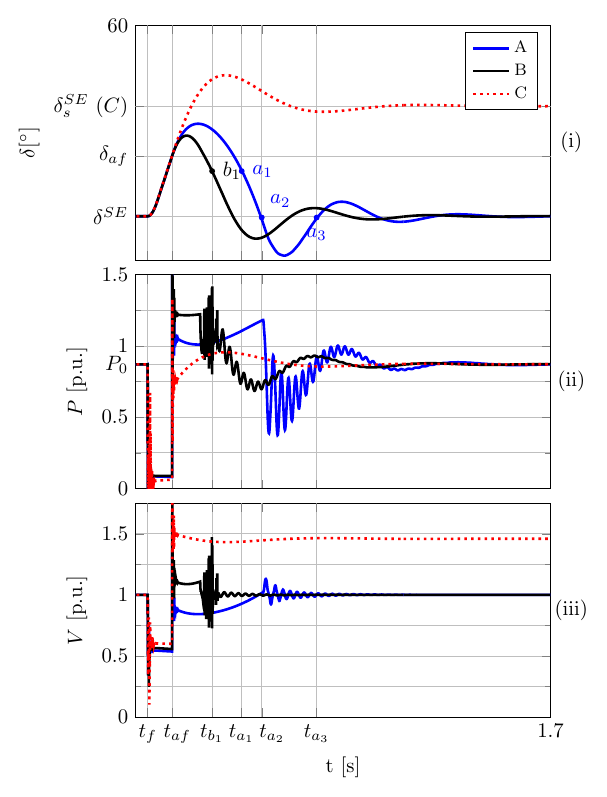}
   \caption{\rrtt{Simulation results for the (i) angles, (ii) powers, and (iii) voltages in Cases A, B, and C.}}
    \label{fig:AnglePowerCaseABC}
\end{figure}

According to Fig.~\ref{fig:CurrentCaseABC}(i), the inverter's output current is limited to $1.2$ p.u. for three Cases A, B, and C. \rrtt{After fault occurance at $t_f=0.05$ s, the active power output drops to near zero. Therefore, the GFM IBR accelerates until clearing fault at $t_{af}=0.15$ s to the post-fault angle of $\delta_{af}=34.93^\circ$.} Among the three post-fault trajectories for these cases, Case B exhibits the fastest convergence to the SEP as illustrated in Fig.~\ref{fig:AnglePowerCaseABC}(i). The GFM IBR converges to the \rt{satSEP} in the Case C because $\delta_s^{SE}(\beta_{\rm C})= 44.52^\circ \in \mathcal{S}$, and the initial point is within the \rt{satSEP's} DOA. Since the GFM IBR is locked in the current-saturation mode, it cannot regulate the \rt{terminal} voltage as shown in Fig.~\ref{fig:AnglePowerCaseABC}(iii). \rt{In Case A, theoretically based on \eqref{eq:Rertun_set}, $\mathcal{R}(\beta_{\rm A}) = [-23.14^\circ, 23.14^\circ]$ and  $\mathcal{S} = [-180^\circ, -32.0455^\circ] \cup [32.0455^\circ, 180^\circ]$. Therefore, after the fault is cleared, it stays in the current-saturation mode until the APC angle becomes less than $23.14^\circ$ at $t=0.517$ s. In other words, the GFM IBR's APC angle is in $\mathcal{S}$ right after the fault is cleared at $t=0.15$ s. It exits this set at $t=0.435 $s, shown as point $a_1$ in Fig.~\ref{fig:CurrentCaseABC}(i) and Fig.~\ref{fig:AnglePowerCaseABC}(i). Since the GFM IBR enters $\overline{\mathcal{R}(\beta_{\rm A}) \cup \mathcal{S}}$, it retains its mode, i.e., the current-saturation mode. It returns to normal operation mode upon entrance to $\mathcal{R}(\beta_{\rm A})-\mathcal{S}$ at $t_{a_2}=0.517 $s, shown as point $a_2$ in Fig.~\ref{fig:CurrentCaseABC}(i) and Fig.~\ref{fig:AnglePowerCaseABC}(i). It enters the $\overline{\mathcal{R}(\beta_{\rm A}) \cup \mathcal{S}}$ again at $t_{a_3}=0.742$ s, shown as $a_3$ in Fig.~\ref{fig:CurrentCaseABC}(i) and Fig.~\ref{fig:AnglePowerCaseABC}(i). It retains its normal operation mode. Since it never again enters $\mathcal{S}$ during its trajectory, it remains in the normal operation mode until it converges to the SEP.} In Case B, $\mathcal{R}(\beta_{\rm B}) = [-45.20^\circ, 45.20^\circ]$. \rt{Therefore, mode oscillations are expected while $\delta \in \mathcal{R}(\beta_{\rm B}) \cap \mathcal{S}= [32.0455^\circ, 45.20^\circ]$. These mode oscillations are reduced using forced saturation introduced in \cite{Arjomandi2024}. The GFM IBR operates as a voltage source from $t_{b_1}=0.314$ s, which is shown as point $b_1$ in Fig.~\ref{fig:CurrentCaseABC}(i) and Fig.~\ref{fig:AnglePowerCaseABC}(i). At this point, the GFM IBR enters $\mathcal{R}(\beta_{\rm B})-\mathcal{S}$.} 

Since $\beta_{\rm B} < \beta_{\rm A} < 0$, more post-fault deceleration is provided in Case B as explained in Section~\ref{sec:TS}. As a result, the angle in Case B converges to the SEP faster than Case A. Therefore, $\beta$ should be properly selected in practice so that (\emph{a}) the GFM IBR does not converge to the \rt{satSEP}, (\emph{b}) more post-fault deceleration is provided.

\begin{figure}[!t]
    \centering
        \includegraphics[width=\linewidth]{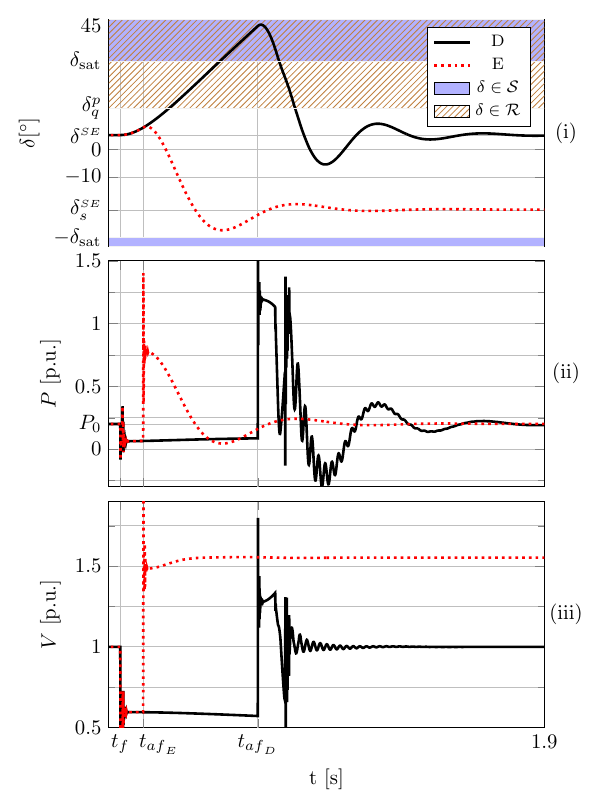}
    \caption{\rrtt{The GFM IBR's (i) angle, (ii) power, and (iii) voltage in Cases D and E.}}
    \label{fig:AnglePowerCaseDE}
\end{figure}
\begin{figure}[!t]
    \centering
  \includegraphics[width=\linewidth]{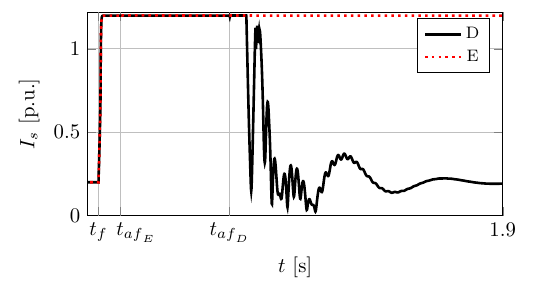}
   \caption{\rrtt{The GFM IBR's current in Cases D and E.}}
    \label{fig:CurrentDE}
\end{figure}
\begin{figure}[!t]
    \centering
    \includegraphics[width=\linewidth]{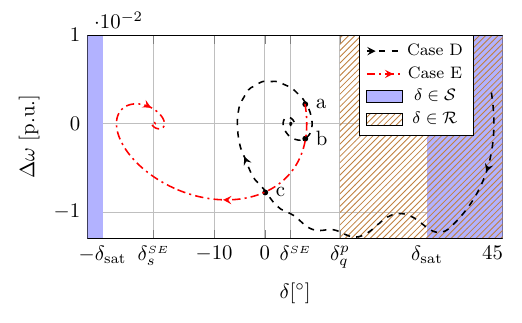}
  \caption{\rrtt{Post-disturbance trajectories of the GFM IBR in Cases D and E where the solid area and hatched area correspond to the set of entering and returning angles, respectively.}}
    \label{fig:TrajrctoriesDE}
\end{figure}

\rrtt{The comparison of cases D and E demonstrates that a short fault duration does not necessarily lead to desired consequences.} The GFM IBR returns to the normal operation mode with some oscillations and converges to the SEP in Case D as shown in Fig.~\ref{fig:AnglePowerCaseDE}(i). However, in Case E, the post-fault angle is out of both $\mathcal{R}(\beta_{\rm E})$ and $\mathcal{S}$. Therefore, it remains in the current-saturation mode as it was during the fault disturbance. Since the power output in the current-saturation mode is more than the power reference at this APC angle, the GFM IBR reduces its angle until it converges to the satSEP, which is out of $\mathcal{S}$, as depicted in Fig.~\ref{fig:AnglePowerCaseDE}(i). According to Fig.~\ref{fig:AnglePowerCaseDE}(ii) in both cases the active power output will be the power reference in the steady-state. However, the goal of voltage regulation is not achieved in Case E in contrast to Case D. Fig.~\ref{fig:CurrentDE} also confirms that Case E remains in the current-saturation mode.

\begin{figure}[t]
    \centering
    \includegraphics[width=\linewidth]{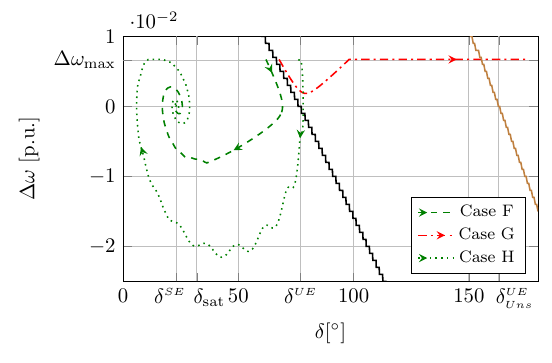}  \caption{Post-disturbance trajectories of the GFM IBR in Cases F, G, and H. Black line and brown line are the borders of DOAs for a GFM equipped and not equipped with CRS; respectively.}
    \label{fig:PhasePortraiteFG}
\end{figure}

\begin{figure}[t]
    \centering
    \includegraphics[width=\linewidth]{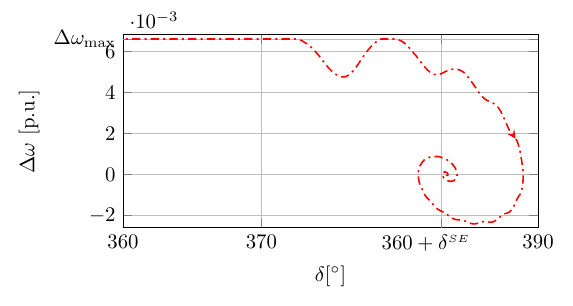}  \caption{\rt{Trajectory of Case G until convergence to the next SEP after passing the unstable equilibrium point 1.}}
    \label{fig:PhasePortraiteG_Comp}
\end{figure}

\begin{figure}[t]
    \centering
    \includegraphics[width=\linewidth]{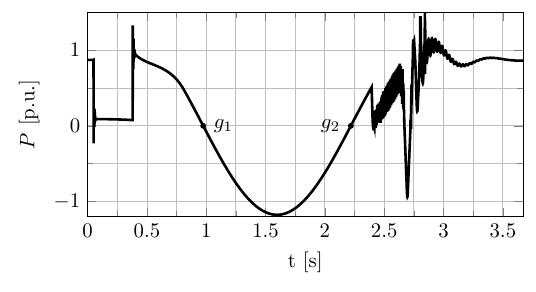}  \caption{\rrtt{The GFM IBR's active power injection in Case G.}}
    \label{fig:PowerCaseGComplete}
\end{figure}

\begin{figure}[t]
    \centering
    \includegraphics[width=\linewidth]{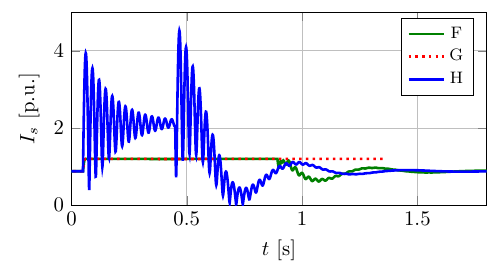}  \caption{\rrtt{The current output of the GFM IBR in Cases F, G, and H.}}
    \label{fig:CurrentFG}
\end{figure}

Notice that in \rt{both} Cases C and E, the GFM IBR converges to the \rt{satSEP}; however, the contributing factors are different. While the \rt{satSEP} was inside $\mathcal{S}$ in Case C (condition $\mathcal{C}_1$ in Section~\ref{sec:TS}), it was out of $\mathcal{S}$ in Case E. Therefore, $\mathcal{C}_1$ cannot be the reason for locking into the current-saturation mode. Indeed, the GFM IBR is locked into the current-saturation mode due to condition $\mathcal{C}_2$. Therefore, it is concluded that if the GFM IBR is being operated in a light loading, the value of $\beta$ should be carefully adjusted so that it is not locked into the current-saturation mode after short disturbances.

The dynamic simulations of Cases F and G confirm that the black line in Fig.~\ref{fig:PhasePortraiteFG}, which is calculated numerically, is the post-fault transient stability border as the black line. This border is much closer to the SEP compared to the border for the unsaturated GFM IBR shown as the brown line in Fig.~\ref{fig:PhasePortraiteFG}.  Fig.~\ref{fig:PhasePortraiteG_Comp} shows that the GFM IBR converges to ($360^\circ+\delta^{SE}$, 0) after passing the unstable equilibrium point 1 in Case G. Although, it is theoretically stable, it is not practically desired. The GFM IBR absorbs active power from \rrtt{$t_{g_1}=0.974$ s} until \rrtt{$t_{g_2}=2.219$ s}. It might cause a considerable over-voltage of the DC link capacitor if its capacitance is not big enough, and as a result, the GFM IBR trips.  Simulation of Case H reveals that even for a post-fault angle larger than those of Case G, being unsaturated helps transient stability. However, Fig.~\ref{fig:CurrentFG} shows that an overcurrent significantly exceeding the current limit appears and lasts for a considerable length of time duration in Case H. Therefore, Case H is not practical and it is simulated here only to confirm current saturation deteriorates transient stability.

\section{Conclusions}
\label{sec:Conclusions}

\rt{The post-disturbance recovery and transient stability analysis for GFM IBRs considering current limitation with constant current angle have been conducted in this paper. These two sets of returning angles and entering angles, which model transition between normal operation mode and current-saturation operation mode have been introduced. The latter only depends on the grid voltage and impedance, and the former depends on the saturated current angle in addition to the grid voltage and impedance and $X/R$ ratio.}

\rt{A GFM IBR may not be able to escape from the current saturation and return to the SEP subject to certain static and dynamic conditions; instead, there is a risk of being locked into a stable point in the current-saturation mode (i.e., satSEP) or even loss of synchronism.  Even a very short fault disturbance can result in the convergence to the satSEP. The saturated current angle plays a significant role in shaping the post-disturbance dynamics of a GFM IBR. This implies the saturated current angle should be carefully regulated. Besides, the traditional CCT or DOA analyses, which are commonly used for transient stability assessment, should be revisited according to such complex behaviors.}

\rt{Although the analyses are only tested and verified under fault disturbances in this paper, they can also be extended for other disturbances that cause deviation of the APC angle from its SEP such as terminal voltage phase jump.}

\section*{Acknowledgments}

The authors would like to thank Prof. Yusheng Xue, Prof.  Damiano Varagnolo, and Dr. Adria Junyent-Ferre for their invaluable feedback and suggestions.

\vfill

\end{document}